\documentclass[aps,showpacs,nofootinbib]{revtex4}

\usepackage{graphicx}
\usepackage{graphics}
\usepackage{amssymb}
\usepackage{amsmath}

\def\slash{\rlap{/}}

\newcommand{\g}{\gamma}

\newcommand{\nslash}{\kern 0.2 em n\kern -0.50em /}
\newcommand{\kslash}{\kern 0.2 em k\kern -0.45em /}
\newcommand{\pslash}{\kern 0.2 em p\kern -0.50em /}
\newcommand{\Sslash}{\kern 0.2 em S\kern -0.50em /}
\newcommand{\Pslash}{\kern 0.2 em P\kern -0.50em /}

\newcommand{\lf}{\left}
\newcommand{\rg}{\right}

\newcommand{\eps}{\epsilon}
\newcommand{\ii}{{\rm i}}

\newcommand{\de}{d}

\newcommand{\mb}{\boldsymbol}

\begin{document}

\title{New observables in longitudinal single-spin asymmetries in semi-inclusive DIS}
\author{Alessandro Bacchetta}
\email{alessandro.bacchetta@physik.uni-r.de}
\affiliation{Institut f{\"u}r Theoretische Physik, Universit{\"a}t Regensburg,
D-93040 Regensburg, Germany}

\author{P.J. Mulders}
\email{mulders@nat.vu.nl}
\affiliation{
Department of Physics and Astronomy, Vrije Universiteit Amsterdam, NL-1081 HV 
Amsterdam, the Netherlands}

\author{Fetze Pijlman}
\email{fetze@nat.vu.nl}
\affiliation{
Department of Physics and Astronomy, Vrije Universiteit Amsterdam, NL-1081 HV 
Amsterdam, the Netherlands}

\begin{abstract}
We analyze longitudinal beam and target
single-spin asymmetries in semi-inclusive deep
inelastic scattering and in jet deep inelastic scattering, including all
possible twist-3
contributions as well as quark mass corrections. We take into account the
path-ordered exponential in the soft correlators and show that it leads to the
introduction of a new distribution and a new fragmentation function
contributing to the asymmetries.

\end{abstract}


\maketitle

%%%%%%%%%%%%%%%%%%%%%%%%%%%%%%%%%%%%%%%%%%%%%%%%%%%%%%%%%%%%%%%%%%%%%
\section{Introduction}

Longitudinal beam and target single-spin asymmetries have been at the center
of the attention lately, since they have been measured by the HERMES and CLAS
experimental
collaborations~\cite{Airapetian:2000tv,Airapetian:2001eg,Airapetian:2002mf,Avakian:2003pk}
and more measurements are planned. 
They were originally believed to be signals of
the so-called T-odd fragmentation functions~\cite{Collins:1993kk}, in particular of the Collins
function~\cite{Efremov:2001cz,Efremov:2001ia,Efremov:2002ut,Ma:2002ns,Efremov:2003eq,Efremov:2003tf,Schweitzer:2003yr}. However, both types of asymmetry can receive contributions also from
T-odd distribution functions~\cite{Sivers:1990cc,Boer:1998nt,Yuan:2003gu,Gamberg:2003pz}, a fact that has often been neglected in
analyses.  
An exhaustive treatment of the contributions of T-odd distribution functions
has not been carried out completely so far, especially up to subleading
order in an expansion in $1/Q$, $Q^2$ being the virtuality of the incident
photon and the only hard scale of the process, and including quark mass
corrections.  It is the purpose of
the present work to describe the longitudinal beam and target spin asymmetries
in a complete way in terms of leading and subleading twist distribution
and fragmentation functions.
We consider both single-particle inclusive DIS, $e + p \to e'+ h+ X$, and
single-jet
inclusive DIS , $e + p \to e'+\rm{jet}+  X$.
We assume factorization holds for these processes, even though at present there
is no factorization proof for 
observables containing 
subleading-twist transverse-momentum dependent functions (only recently proofs
for the leading-twist case have been presented in Refs.~\cite{Ji:2004wu,Ji:2004xq}).

We devote particular attention to the claims presented in
Ref.~\cite{Goeke:2003az}, where it was suggested that the decomposition of the
quark correlator should contain more terms than the ones considered in
Refs.~\cite{Mulders:1996dh,Boer:1998nt}. The inclusion of the gauge link in
the proper definition of the correlator, in fact, introduces a dependence on
the light cone vector, $n_-$, that defines the direction along which the
path-ordered exponential is running. The inclusion of this new degree of
freedom spoils the Lorentz-invariance relations among distribution
functions as pointed out in Ref.~\cite{Mulders:1996dh,Boer:1998nt}, 
but the study in Ref.~\cite{Goeke:2003az} is incomplete as 
an extra term in the decomposition of the unpolarized correlator has
been neglected. This gives rise to a {\em new} distribution function and a
{\em new} fragmentation function. 
We take
these new terms into account and study their effect on the
longitudinal asymmetries. Evidence -- either from experiments or from model
calculations -- for the existence of these new functions could support the
necessity of introducing the gauge-link direction in the decomposition of the
correlator.

%%%%%%%%%%%%%%%%%%%%%%%%%%%%%%%%%%%%%%%%%%%%%%%%%%%%%%%%%%%%%%%%%%%%%
\section{Unpolarized target}

We adopt the point of view of Ref.~\cite{Goeke:2003az}
and complete the treatment presented there. 
We introduce first of all the four-momentum of the target, $P$, and that of 
the quark, $p$, and their decomposition in terms of light-cone
vectors 
\begin{align} 
P^\mu &= P^+ n_{+}^{\mu} + \frac{M^2}{2 P^+}n_{-}^{\mu}, &
p^\mu &= x P^+ n_{+}^{\mu} + p^- n_{-}^{\mu} + p_{T}^{\mu}.
\end{align}

To construct the hadronic tensor and consequently the cross sections, we start
from the distribution correlation function (for the moment being we shall
consider the target to be unpolarized)
\begin{equation}  
\Phi^{[+]}
%_{\mathrm{unpol}}
(x,p_T)=\int \de p^- \Phi^{[+]}(P,p,n_-)
\end{equation}
where $\Phi^{[+]}
%_{ij}_{\mathrm{unp}}
(x,p_T)$ includes the transverse link \cite{Ji:2002aa,Belitsky:2002sm}
\begin{equation} 
{\Phi^{[+]}_{ij}}
%_{\mathrm{unpol}}
(x,p_T)= \int
        \frac{\mathrm{d} \xi^-\ \mathrm{d}^2 \xi_T}{(2\pi)^{3}}\;
 e^{+\ii p \cdot \xi}
       \langle P|\bar{\psi}_j(0)\,{\cal L}_{[0^-, \infty^-]}
{\cal L}_{[0_T, \xi_T]}{\cal
L}_{[\infty^-, \xi^-]}\psi_i(\xi)|P \rangle \bigg|_{\xi^+=0}\,,
\label{e:phi}
 \end{equation}  
The notation ${\cal L}_{[a,b]}$ indicates a straight gauge link running 
from $a$ to $b$.

The most general form of the correlation function $\Phi^{[+]}$ complying with
Hermiticity and parity constraints reads 
\begin{equation} \begin{split} 
 \label{eq:phiunpol1}
\Phi^{[+]}(P,p,n_-) &= M A_{1} + \Pslash A_{2} + \pslash A_{3} + \frac{\ii}{2
  M}\bigl[\Pslash, \pslash \bigr] A_{4} 
\\ & \quad
       + \frac{M^2}{P\cdot n_-}\nslash_- B_1 + \frac{\ii M}{2P\cdot n_-}
        \bigl[\Pslash, \nslash_- \bigr] B_{2} + \frac{\ii M}{2P\cdot n_-}
        \bigl[\pslash, \nslash_- \bigr] B_{3}
\\ & \quad
        + \frac{1}{P\cdot n_-}\g_5 \eps^{\mu \nu \rho \sigma}\g_\mu P_\nu n_{-
          \rho} p_{\sigma} B_4.
\end{split} \end{equation}  
The last term was neglected in Ref.~\cite{Goeke:2003az}. It is a T-odd and
chiral-even structure.

%and we keep only the leading and subleading terms in $1/P^+$
%\begin{equation} \begin{split} 
% \label{eq:phiunpol2}
%\Phi^{[+]}(P,p;n_-) &\approx P^+ \lf(A_{2} + x A_{3} \rg) \nslash_+ + P^+
%\frac{\ii}{2M} \bigl[\nslash_+, \pslash_T \bigr] A_{4} + M A_{1} + \pslash_T
%A_{3} 
%\\ & \quad 
%+ \lf[\lf(\frac{P^+ p^-}{M}-\frac{x M}{2}\rg)A_{4} + M B_{2} + x M 
%B_{3}\rg]\frac{\ii}{2} \bigl[\nslash_+, \nslash_- \bigr]
%+ \g_5 \eps_T^{\rho \sigma} \g_{\rho} p_{T \sigma} B_4
%\end{split} \end{equation}  
%leading to

Keeping only the leading and subleading terms in $1/P^+$ we obtain
\begin{equation} \begin{split} 
 \label{eq:phiunpol3}
\Phi^{[+]}(x, p_T) & \equiv \int \de p^- \Phi^{[+]}(P,p;n_-) \\
& = \frac{1}{2} \, \biggl\{ f_1 \nslash_+ +\ii h_1^\perp \frac{ \bigl[
  \pslash_T, \nslash_+ \bigr]}{2M}\biggr\} \\
& \quad + \frac{M}{2 P^+}\,\biggl\{ e +f^\perp \frac{\pslash_T}{M}+ \ii h
\frac{ \bigl[\nslash_+, \nslash_- \bigr]}{2} + g^\perp
\g_5\,\frac{\eps_T^{\rho \sigma} \g_{\rho} p_{T \sigma}}{M}\biggr\}, 
\end{split} \end{equation}  
where the new function $g^\perp$ was introduced.
The functions on the right-hand side depend on $x$ and $p_T^2$ and they are
explicitly
\begin{align} 
f_1(x, p_T^2) & = 2 P^+ \int \de p^- \lf(A_{2} + x A_{3}\rg),
&
h_1^{\perp}(x, p_T^2) & = 2 P^+ \int \de p^- \lf(-A_{4}\rg), \nonumber
\\
e(x, p_T^2) & = 2 P^+ \int \de p^- A_{1},
&
f^{\perp}(x, p_T^2) & = 2 P^+ \int \de p^- A_{3}, \nonumber
\\
h(x, p_T^2) & = 2 P^+ \int \de p^- \lf(\frac{p\cdot P -x M^2}{M^2}A_{4}+
B_{2}+ x B_{3} \rg),
&
g^{\perp}(x, p_T^2) & = 2 P^+ \int \de p^- B_{4}. \nonumber
\end{align} 
The last function has never been discussed in the literature so far, but it 
could
correspond to the object calculated in the framework of the diquark model in
Refs.~\cite{Afanasev:2003ze} and \cite{Metz:2004je}, as we shall see after we
study the expression for the asymmetry.

The structure of the fragmentation correlator $\Delta$ is analogous to that of
$\Phi$, including in particular the presence of a new fragmentation function
$G^{\perp}$. The complete expression up to subleading twist is
\begin{equation} \begin{split} 
 \label{eq:delta}
\Delta^{[-]}(z, k_T) & \equiv \int \de k^+ \Delta^{[-]}(P_h,k;n_+) \\
& = z \, \biggl\{ D_1 \nslash_- + \ii H_1^\perp \frac{ \bigl[
  \kslash_T, \nslash_- \bigr]}{2M_h}\biggr\} \\
& \quad + \frac{z M_h}{P_h^-}\,\biggl\{ E +D^\perp \frac{\kslash_T}{M_h}+ \ii H
\frac{ \bigl[\nslash_-, \nslash_+ \bigr]}{2} + G^\perp
\g_5\,\frac{\eps_T^{\rho \sigma} \g_{\rho} k_{T \sigma}}{M_h}\biggr\}. 
\end{split} \end{equation}  

The transverse gauge link leads to full
color gauge invariant expressions at leading and next-to-leading 
order for the hadronic tensor. 
%In this section we will express this
%tensor in terms of the unpolarized distribution and fragmentation functions
%including $h_1^\perp$ and $h$.
%
The tree level result at leading and next-to-leading order was given
by Ref.~\cite{Mulders:1996dh}, Eq.~(73). In that paper the need to consider 
transverse
gluon fields at infinity was mentioned but the 
transverse gauge link was not taken into account. Taking this link into
account, which allows T-odd distribution functions including $g^\perp$,
does not change the procedure of obtaining the hadronic tensor (compare
the expressions for the hadronic tensor given in Ref.~\cite{Boer:2003cm} with 
the ones in
Ref.~\cite{Mulders:1996dh}). The main difference is the inclusion of the new 
distribution and fragmentation
functions. We obtain (using a notation similar to that 
of Ref.~\cite{Boer:2003cm})
\begin{equation} \begin{split} 
2M W^{\mu\nu} &= \int 
%\mathrm{d} p^-\ \mathrm{d} k^+\ 
\mathrm{d}^2 p_T\
\mathrm{d}^2 k_T\ \delta^2( \boldsymbol{p}_T + \boldsymbol{q}_T - 
\boldsymbol{k}_T ) 
\mathrm{Tr} \biggl[ \Phi^{[+]} 
%(p)
(x,p_T)
\gamma^\mu \Delta^{[-]} 
%(k)
(z,k_T)
\gamma^\nu  
\\
& \quad - \gamma_\alpha \frac{\slash{n}_+}{Q\sqrt{2}} \gamma^\nu
{\Phi^{[+]}_{ \partial^{-1}G}}^{\alpha} 
%(p)
(x,p_T)
\gamma^\mu \Delta 
%(k)
(z,k_T) - \gamma^\alpha \frac{\slash{n}_-}{Q\sqrt{2}} \gamma^\mu
{\Delta^{[-]}_{\partial^{-1}G}}^{\alpha} 
%(k)
(z,k_T) \gamma^\nu \Phi^{[+]} 
%(p)
(x,p_T) %\nonumber \\
%& &
+ (\mu \leftrightarrow
\nu)^* \biggr] \label{eq1},
\end{split} \end{equation}   
where the $(\mu \leftrightarrow \nu)^*$ acts on the last two terms only and
\begin{align}
\lf({\Phi^{[\pm]}_{\partial^{-1}G}}\rg)_{ij}^{\alpha} 
%(p)
(x,p_T) 
&=\int \mathrm{d} p^- 
\int \frac{\mathrm{d}^4 \xi}{(2\pi)^4} 
e^{ip\xi} \langle P,S| \overline{\psi}_j(0) \int_{\pm \infty}^{\xi^-}
 \mathrm{d}\eta^-\ U^{[\pm]}(0,\eta)
G^{+\alpha}(\eta) U^{[\pm]}(\eta,\xi) \psi_i(\xi) |P,S \rangle
 \bigg|_{\begin{array}{l}
\eta^+ = \xi^+ \\ \eta_T = \xi_T \end{array}},\\
{\Phi^{[+]}_{\partial^{-1}G}}^{\alpha} 
%(p)
(x,p_T) &= {\Phi^{[+]}_D}^\alpha  
%(p)
(x,p_T) - 
{\Phi^{[+]}_\partial}^\alpha  
%(p)
(x,p_T),\\
{\Delta^{[-]}_{\partial^{-1}G}}^{\alpha} 
%(k)
(z,k_T) &= {\Delta^{[-]}_D}^\alpha  
%(k)
(z,k_T) - 
{\Delta^{[-]}_\partial}^\alpha  
%(k)
(z,k_T).
\end{align}
Note that in the derivation of the last two equations we made use of identities
which also relates the Qiu-Sterman mechanism to the Sivers 
effect~\cite{Boer:2003cm,Boer:2003xz,Ma:2003ut}.

Certain traces of correlation functions which contain a covariant derivative 
can be related to
distribution and fragmentation functions by using the equations of motion. 
Including T-odd and longitudinal target polarization we obtain for the
distribution functions
\begin{align}
\frac{1}{2} \mathrm{Tr} \big[ {\Phi^{[+]}_{\partial^{-1}G}}^{\alpha} 
 \sigma_\alpha^{\ +} 
\big]
&= i \bigl( M x\, e - m\, f_1 - i M x\, h \bigr) - \frac{p_T^2}{M}\,
h_1^\perp,\\
\frac{1}{2} \mathrm{Tr} 
\big[ {\Phi^{[+]}_{\partial^{-1}G}}^{\alpha} 
 \, i \sigma_\alpha^{\ +} \gamma_5 \big] &=
-m S_L \,g_{1L} + i M x S_L\, e_L +
M x S_L\, h_L - \frac{p_T^2}{M} S_L\, h_{1L}^\perp,\\
\begin{split}
\frac{1}{2} \mathrm{Tr} \big[ {\Phi^{[+]}_{\partial^{-1}G}}^{\alpha} \,
 \gamma^+ \big] &=
\frac{1}{2}\, i \epsilon_T^{\alpha\beta} \mathrm{Tr} \big[ {\Phi^{[+]}_{\partial^{-1}G}}_\beta \,
\gamma^+ \gamma_5 \big] 
+
p_T^\alpha \bigl( x \,f^\perp + 
i \frac{m}{M}\, h_1^\perp + i x\, g^\perp -  f_1 \bigr) 
\\
&\quad - \epsilon_T^{\alpha\beta} p_{T\beta} \bigl(
x S_L \,f_L^\perp - i \frac{m}{M} S_L\, h_{1L}^\perp + ixS_L\, g_L^\perp
-i S_L \,g_{1L} \bigr).
\end{split}
%\\
%\frac{1}{4}\mathrm{Tr} \big[ {\Phi^{[+]}_{\partial^{-1}G}}^\alpha 
%\gamma^+ \gamma_5 \big]
%& = \frac{p_{T}^{\alpha}}{2} \bigl(
%x S_L f_L^\perp - i \frac{m}{M} S_L h_{1L}^\perp + ixS_L g_L^\perp
%-i S_L g_{1L} \bigr)
\end{align}
Using these identities we can calculate the hadronic tensor in
Eq.~(\ref{eq1}) by using FORM~\cite{Vermaseren:2000nd}. 
We obtain the unpolarized parts of Eq.\ (77) and Eq.\ (78) of
Mulders and Tangerman~\cite{Mulders:1996dh} (denoted by 
$2MW^{\mathrm{[MT]}\mu\nu}_{\mathrm{U}}$) together with some extra terms
\begin{equation} \begin{split} 
2MW^{\mu\nu}_{\mathrm{U}} &= 2MW^{\mathrm{[MT]}\mu\nu}_{\mathrm{U}}
 + 
 2 z_h \int \mathrm{d}^2 p_T\ \mathrm{d}^2 k_T\ 
\delta^2(\boldsymbol{p}_T + \boldsymbol{q}_T - \boldsymbol{k}_T) 
\\
&\quad \times \biggl\{ -\left(g_\perp^{\mu\nu} \boldsymbol{k}_\perp \cdot 
\boldsymbol{p}_\perp + 
k_\perp^{\{ \mu} p_\perp^{\nu\}} \right) 
\frac{1}{M M_h}h_1^\perp H_1^\perp  %\nonumber\\
%&&
+p_\perp^{\{ \mu} t^{\nu \}} 
\frac{2 \boldsymbol{k}_T^2 }{M M_h Q}h_1^\perp H_1^\perp
%\nonumber 
\\
& \quad
+p_\perp^{\{ \mu} t^{\nu \}} \frac{2 M_h }{z_h M Q}h_1^\perp H
%\nonumber\\
%&& 
+k_\perp^{\{ \mu} t^{\nu \}} \frac{2 x M }{M_h Q}h H_1^\perp
%\nonumber\\
%&& 
- t^{[\mu } p_\perp^{\nu ]} \frac{2 i m }{MQ} h_1^\perp D_1
%\nonumber 
\\
&\quad
+t^{[\mu } p_\perp^{\nu ]} \frac{2 i M_h }{z_hMQ}h_1^\perp E
-t^{[\mu } p_\perp^{\nu ]} \frac{2 i}{Q} x g^\perp D_1
-t^{[\mu } k_\perp^{\nu ]} \frac{2 i}{Q}f_1 \frac{G^\perp}{z}
\biggr\}.
\end{split} \end{equation}  
Notice that the hadronic tensor we obtain is electromagnetic gauge invariant
($q_{\mu} W^{\mu \nu} = 0$). Gauge invariance is insured thanks to the
contribution of the quark-gluon-quark correlator $\Phi_{\partial^{-1}G}$. 
%This contribution
%has been neglected in
%the model calculation of Ref.~\cite{Afanasev:2003ze} and this probably
%is the reason why that calculation is not gauge invariant. On the other hand, 
%our formalism seems
%not to include the extra contributions explored in Ref.~\cite{Metz:2004je},
%which then seem not to be essential to restore gauge invariance.

Unpolarized T-odd distribution functions can be measured for instance 
in beam single-spin
asymmetries. The polarization of the beam forms an antisymmetric structure
that has to be contracted with the antisymmetric part 
of $W^{\mu\nu}$. This part consists of either T-odd
distribution functions with T-even fragmentation functions or vice versa.

We find that the $A_{LU}$ asymmetry is given by\footnote{We use the same definition
of azimuthal angles as in Ref.~\cite{Mulders:1996dh}}
\begin{equation}
A_{LU} = \frac{\big(L_{\mu\nu}^{\lambda_e=1}-L_{\mu\nu}^{\lambda_e=-1}\big)
               2M W^{\mu\nu}_{\mathrm{U}}}
              {\int \mathrm{d}^2 P_h^\perp \ 
              \big(L_{\mu\nu}^{\lambda_e=1}+L_{\mu\nu}^{\lambda_e=-1}\big)
              2MW^{\mu\nu}_{\mathrm{U}}} = \frac{2
%z
               y \sqrt{1-y}}{(1-y + y^2/2)\, f_1 
D_1}\,\sin{\phi_h}\frac{M}{Q}\, {\cal A},
\end{equation}
where
\begin{equation} \begin{split} 
{\cal A} &= {\cal I} \biggl\{\frac{\hat{\boldsymbol{h}}\cdot 
\boldsymbol{k}_{\perp}}{M_h}
 \lf[\lf(x\, e - \frac{m}{M}\, f_1\rg) H_1^{\perp} 
+\frac{M_h}{M}\,f_1\,\frac{G^\perp}{z}\rg]- \frac{\hat{\boldsymbol{h}}\cdot 
\boldsymbol{p}_{\perp}}{M}\lf[
\frac{M_h}{M}\, h_1^{\perp}\lf(\frac{E}{z}-\frac{m}{M_h}\, D_1\rg)-x\, g^\perp 
D_1\rg] \biggr\} \end{split}
\end{equation}
Here we introduced the symbol 
$\hat{\boldsymbol{h}}=\boldsymbol{P}_{h \perp}/|\boldsymbol{P}_{h
  \perp}|$ and the
shorthand notation
\begin{equation} \begin{split} 
{\cal I}\bigl\{\dotsb \bigr\} & \equiv
 \int \de^2 \! \mb{p}_T \de^2 \mb{k}_T \,\delta^{(2)}\Bigl(\mb{p}_T
        -\frac{\mb{P}_{h \perp}}{z}-\mb{k}_T\Bigr) \;\bigl\{\dotsb \bigr\}
\label{e:convolution}
\end{split} \end{equation}  
Preliminary measurements of this asymmetry have been presented by
the CLAS and HERMES collaboration~\cite{Avakian:2003pk,Avetisyan:2004}. The
interpretation of such asymmetry has to take into account the possible
contribution of $g^{\perp}$ and  $G^{\perp}$.

From now on we will avoid writing explicitly the charge weighted 
summation over the quark flavors and omit the flavor indices of the functions.

To deconvolute the $\sin \phi_h$ asymmetry 
in a clean manner, it is necessary to introduce a unit vector 
$\hat{\boldsymbol{a}}$ (fixed with respect to the lepton scattering plane),
weight the asymmetry with 
$\boldsymbol{P}_{h\perp} \cdot \hat{\boldsymbol{a}}$ and integrate over 
$P_{h\perp}$. Defining
\begin{equation}
A_{\ldots}^{\boldsymbol{P}_{h\perp} \cdot \hat{\boldsymbol{a}}} = 
\int \mathrm{d}^2 P_{h\perp}\; \boldsymbol{P}_{h\perp} \cdot \hat{\boldsymbol{a}} 
\ A_{\ldots},
\end{equation}
we find that
\begin{equation} \begin{split}
A_{LU}^{\boldsymbol{P}_h^\perp \cdot \hat{\boldsymbol{a}}} = 
\frac{2y\sqrt{1-y}}{(1-y+y^2/2)\, f_1 D_1} \sin \phi_a \frac{M M_h}{Q} 
\bigg[ & \frac{m}{M}z\, f_1 H_1^{\perp(1)} - \frac{M_h}{M}\, f_1 G^{\perp(1)} 
- xz\, e H_1^{\perp(1)} \\
&+ \frac{m}{M_h}z\, h_1^{\perp(1)} D_1 -  h_1^{\perp(1)} E
+ \frac{ M}{M_h}xz\, g^{\perp(1)} D_1 \bigg]. \end{split}
\end{equation}
The distribution and fragmentation functions on the right-hand side depend
only on $x$ and $z$ respectively. 
The asymmetry is maximized by choosing $\hat{\boldsymbol{a}}$ perpendicular to
the lepton scattering plane, one obtains ($\phi_a = \pi / 2$)
\begin{equation}
\begin{split}
A_{LU}^{|\boldsymbol{P}_h^\perp | \sin \phi_h} = 
\frac{2y\sqrt{1-y}}{(1-y+y^2/2)\, f_1 D_1}\ \frac{M M_h}{Q} 
\bigg[ & \frac{m}{M}z\, f_1 H_1^{\perp(1)} - \frac{M_h}{M}\, f_1 G^{\perp(1)} 
- xz\, e H_1^{\perp(1)} \\
&+ \frac{m}{M_h}z\, h_1^{\perp(1)} D_1 -  h_1^{\perp(1)} E
+ \frac{M}{M_h}xz\, g^{\perp(1)} D_1 \bigg]. \end{split}
\label{e:Aluw}
\end{equation}
Apart from the
presence of $g^{\perp}$, $G^{\perp}$, the terms with quark masses, and a factor 2 difference in the
definition, the expression for the weighted asymmetry corresponds
to Eq.~(21) of Ref.~\cite{Gamberg:2003pz} (the different sign is due to a
different definition of the azimuthal angle). A similar result was also
obtained in Ref.~\cite{Yuan:2003gu}.

In jet semi-inclusive DIS with massless quarks 
$H_1^{\perp}$, $E$ and $G^\perp$ vanish and $D_1$
reduces
to $\delta(1-z_h)$. The asymmetries are in that case directly proportional to the
T-odd distribution function $g^{\perp}$.
\begin{equation}
A_{LU,j}^{|\boldsymbol{P}_{h\perp} | \sin \phi_h} =  \frac{M^2}{Q}
\frac{2 y \sqrt{1-y}}{(1-y+y^2/2)}\ 
\frac{x\,g^{\perp(1)}}{f_1}
\end{equation}
In Ref.~\cite{Afanasev:2003ze} and
\cite{Metz:2004je} model calculations of this jet asymmetry have been
studied.  Without the introduction of the function $g^{\perp}$ this
asymmetry would vanish, therefore suggesting a connection between the model
calculations of the asymmetry and the function $g^{\perp}$.
An experimental study of this asymmetry in jet semi-inclusive DIS (e.g. at
ZEUS, H1 or at a future facility as eRHIC)  would be
important to establish if the functions  $g^{\perp}$ exists. 
Its measurement
would allow also a cleaner study of the terms containing the functions $e$ and
$h_1^{\perp}$ in the asymmetry of Eq.~\ref{e:Aluw}. Note that perturbative contributions
to this asymmetry have also to be taken into account~\cite{Hagiwara:1983cq,Ahmed:1999ix}.

%%%%%%%%%%%%%%%%%%%%%%%%%%%%%%%%%%%%%%%%%%%%%%%%%%%%%%%%%%%%%%%%%%%%%
\section{Target polarized along the virtual photon}

So far, no complete study has been performed including the T-odd distribution
function $f_L^{\perp}$ and
$G^{\perp}$.
When taking longitudinal target polarization into account, 
the use of the vector $n_-$ in this case generates no other structures
than the ones already presented in Ref.~\cite{Boer:1998nt}, 
even
though it changes the relation between the distribution functions and the
amplitudes, invalidating Lorentz invariance relations. 

We find that the
longitudinal polarized parts of Eq.\ (77) and Eq.\ (78) of
Ref.~\cite{Mulders:1996dh} (denoted by 
$2MW^{\mathrm{[MT]}\mu\nu}_{\mathrm{L}}$) 
\begin{equation} \begin{split} 
2MW^{\mu\nu}_{\mathrm{L}} &= 2MW^{\mathrm{[MT]}\mu\nu}_{\mathrm{L}} \\
&\quad - \frac{4}{Q} S_L \epsilon_\perp^{\rho \{ \mu} k_{\perp\rho} t^{\nu \} }
g_{1L} G^\perp 
+ \frac{4xz}{Q} S_L \epsilon_\perp^{\rho \{ \mu} p_{\perp\rho} t^{\nu \} }
f_L^\perp D_1 + \frac{4i xz M}{M_h Q} S_L \epsilon_\perp^{\rho [ \nu} k_{\perp\rho}
t^{\mu ]} e_L H_1^\perp.
\end{split} \end{equation}  

The asymmetry reads
\begin{eqnarray}
A_{UL} &=& \frac{L_{\mu\nu}^{\mathrm{U}} \big(2M W^{\mu\nu}_{S_L=1} -
                                            2M W^{\mu\nu}_{S_L=-1}\big)}
              {\int \mathrm{d}^2 P_h^\perp\ L_{\mu\nu}^{\mathrm{U}} 
              \big(2M W^{\mu\nu}_{S_L=1} +
                                            2M W^{\mu\nu}_{S_L=-1}\big)
                                            }
                                            \nonumber\\
&=&\frac{1}{(1-y + y^2/2)\, f_1 D_1}\,\Bigl[(1-y)\,\sin{2 
\phi_h}\,{\cal B}
+ 2\,(2-y)\sqrt{1-y}\,\frac{M}{Q}\, \sin{\phi_h}\,{\cal C}\Bigl],
\end{eqnarray}
where
\begin{align} 
{\cal B} &= {\cal I} \biggl\{\frac{2\,\hat{\boldsymbol{h}}\cdot
  \boldsymbol{k}_{\perp}\,\hat{\boldsymbol{h}}\cdot
  \boldsymbol{p}_{\perp}-\boldsymbol{k}_{\perp}\cdot \boldsymbol{p}_{\perp}}{M M_h}\,h_{1L}^{\perp}\,H_{1}^{\perp}\biggr\}
\\
{\cal C} &=
{\cal I} \biggl\{\frac{\hat{\boldsymbol{h}}\cdot \boldsymbol{k}_{\perp}}{M_h}
 \lf[\lf(x\, h_L - \frac{m}{M}\, g_{1L}\rg) H_1^{\perp} 
+\frac{M_h}{M}\,g_{1L}\,\frac{G^\perp}{z}\rg]+ \frac{\hat{\boldsymbol{h}}\cdot 
\boldsymbol{p}_{\perp}}{M}\lf[
\frac{M_h}{M}\, h_{1L}^{\perp} \frac{\tilde{H}}{z} -x\, f_{L}^\perp D_1\rg] \biggr\}
\end{align}
where we introduced the function 
$\tilde{H}= H + H_1^{\perp} z {\boldsymbol{k}}_{\perp}^2 / M_h^2  $.

Following the same steps as described
in the previous section to deconvolute the $\sin \phi_h$ asymmetry, we find
\begin{equation} \begin{split}
A_{UL}^{\boldsymbol{P}_{h\perp} \cdot \hat{\boldsymbol{a}}} = 
\frac{2(2-y)\sqrt{1-y}}{(1-y+y^2/2) f_1 D_1} \sin \phi_a
\frac{M M_h}{Q} \bigg[ & 
\frac{m}{M}z\ g_1 H_1^{\perp(1)} - \frac{M_h}{M} g_1 G^{\perp(1)} - 
xz\ h_L H_1^{\perp(1)} \\
& + h_{1L}^{\perp(1)} \tilde{H} - \frac{M}{M_h}xz\ f_L^{\perp(1)} D_1 \bigg]. 
\end{split}
\end{equation}
Again, the asymmetry is maximized by choosing $\phi_a = \pi/2$. For this 
particular $\hat{\boldsymbol{a}}$  the weight $
\boldsymbol{P}_{h\perp} \cdot \hat{\boldsymbol{a}} $ reduces to 
$| \boldsymbol{P}_{h\perp} | \sin \phi_h$. Neglecting quark masses, 
the $\sin \phi_h$ asymmetry for jet production 
reduces to
\begin{equation}
A_{UL,j}^{|\boldsymbol{P}_{h\perp} |\sin \phi_h} = -\frac{M^2}{Q}\,
\frac{2(2-y)\sqrt{1-y}}{(1-y+y^2/2)}\
\frac{x\, f_L^{\perp(1)}}{f_1}
\end{equation} 
This is the situation studied in the
model calculations of Ref.~\cite{Metz:2004je}.

To deconvolute the $\sin 2\phi_h$ term we introduce a new unit vector 
$\hat{\boldsymbol{b}}$ and weight with 
$\boldsymbol{P}_{h\perp} \cdot \hat{\boldsymbol{a}}\;
\boldsymbol{P}_{h\perp} \cdot \hat{\boldsymbol{b}}$. We
obtain
\begin{equation}
A_{UL}^{\boldsymbol{P}_{h\perp} \cdot \hat{\boldsymbol{a}}\;
        \boldsymbol{P}_{h\perp} \cdot \hat{\boldsymbol{b}}} = M M_h
\frac{2 z^2 (1-y) \sin (\phi_a + \phi_b )}{(1-y+y^2/2) f_1 D_1} 
h_{1L}^{\perp(1)} H_1^{\perp(1)}.
\end{equation}
Choosing $\hat{\boldsymbol{a}}$ perpendicular and $\hat{\boldsymbol{b}}$ tangent
to the lepton scattering plane, one finds the maximal 
asymmetry
\begin{equation}
A_{UL}^{\boldsymbol{P}_{h\perp}^2 \sin (2\phi_h)} = M M_h
\frac{4 z^2 (1-y)}{(1-y+y^2/2) f_1 D_1} 
h_{1L}^{\perp(1)} H_1^{\perp(1)}.
\end{equation}

%%%%%%%%%%%%%%%%%%%%%%%%%%%%%%%%%%%%%%%%%%%%%%%%%%%%%%%%%%%%%%%%%%%%%
\section{Target polarized along the beam}

In experiments the target is
polarized along the beam direction and not along 
the virtual photon direction (we
will denote the longitudinal polarization along the beam as $L'$ to
distinguish it from that along the virtual photon, $L$). To
write the complete $UL'$ asymmetry, therefore, we should include also the 
leading twist part of the $UT$ asymmetry, which appears with a $1/Q$ 
suppression~\cite{Korotkov:2001jx}.
When dealing with the $UT$ asymmetry, we have to check whether the
introduction of the $n_-$ vector in the parameterization of the correlator
generates new structures or not. It turns out that some new structures appear
at subleading twist, and one new structure appears also at leading
twist: it is the T-even and chiral-odd structure $[\pslash_T, \nslash_+]
\eps_T^{\rho \sigma} p_{\rho} S_{\sigma}$. However, for the $A_{UL'}$ asymmetry
this new term is
indistinguishable from the transversity and it can absorbed into its
definition, leading to no extra distribution functions.

The final answer is
\begin{equation}
\begin{split}
A_{UL'} &= \frac{L_{\mu\nu}^{\mathrm{U}} \big( 2MW^{\mu\nu}_{S_L'=1} - 
                                               2MW^{\mu\nu}_{S_L'=-1}\big)}
                {\int \mathrm{d}^2 P_{h\perp}\ L_{\mu\nu}^{\mathrm{U}}
                 \big( 2MW^{\mu\nu}_{S_L'=1} + 
                                               2MW^{\mu\nu}_{S_L'=-1}\big)
                                           }
                                            \\                               
&=\frac{1}{(1-y + y^2/2)\, f_1 D_1}\,\biggl\{(1-y)\,\sin{2 
\phi_h}\,{\cal B}
\\
&\quad+ 2\sqrt{1-y}\frac{M}{Q} \Bigl[\sin{\phi_h}\bigl( (2-y)\, {\cal C}-(1-y)\, 
{\cal D}-(1-y + y^2/2)\,{\cal E}\bigr)-\sin{3 \phi_h}\,(1-y) \,{\cal 
F}\biggr]\biggr\},
\end{split}
\end{equation}
where
\begin{align} 
{\cal D} &= {\cal I} \biggl\{\frac{\hat{\boldsymbol{h}}\cdot \boldsymbol{k}_{\perp}}{M_h}\,x\, h_1\, 
H_1^{\perp} \biggr\}
\\
{\cal E} &= {\cal I} \biggl\{\frac{\hat{\boldsymbol{h}}\cdot \boldsymbol{p}_{\perp}}{M}\, x\,
f_{1T}^{\perp}\, D_1 \biggr\}
\\
{\cal F} &={\cal I} \biggl\{\frac{4\,(\hat{\boldsymbol{h}}\cdot
  \boldsymbol{p}_{\perp})^2 \hat{\boldsymbol{h}}\cdot
  \boldsymbol{k}_{\perp}-2\,\hat{\boldsymbol{h}}\cdot
  \boldsymbol{p}_{\perp}\,\boldsymbol{k}_{\perp}\cdot 
  \boldsymbol{p}_{\perp}-\boldsymbol{p}_{\perp}^2\,\hat{\boldsymbol{h}}\cdot
  \boldsymbol{k}_{\perp}}{2 M^2 M_h}\, x\ h_{1T}^{\perp}\,H_{1}^{\perp}\biggr\}
\end{align}
This asymmetry has been measured by the HERMES
collaboration~\cite{Airapetian:2000tv,Airapetian:2001eg,Airapetian:2002mf}.
However, none of the interpretations given so far takes into account the
contribution of the function $f_L^\perp$ and $G^\perp$ in ${\cal C}$, while 
only a few
discuss the contribution of the Sivers function $f_{1T}^{\perp}$ in  ${\cal E}$
  \cite{Efremov:2003tf,Schweitzer:2003yr}.

The single and double weighted asymmetries read
\begin{align}
\begin{split}
A_{UL'}^{\boldsymbol{P}_{h\perp} \cdot \hat{\boldsymbol{a}}} &= 
A_{UL}^{\boldsymbol{P}_{h\perp} \cdot \hat{\boldsymbol{a}}} +
\frac{2\sqrt{1-y}}{(1-y+y^2/2)\, f_1 D_1} \sin \phi_a \frac{M M_h}{Q}
\bigg[
\begin{aligned}[t]
&( 1-y ) \bigg( xz h_{1} H_1^{\perp(1)} \bigg) \\
&-( 1-y+y^2/2 ) \bigg( \frac{M}{M_h}xz\ f_{1T}^{\perp(1)} D_1 \bigg) 
\bigg] 
\end{aligned}
\end{split}
\\
 A_{UL'}^{\boldsymbol{P}_{h\perp} \cdot \hat{\boldsymbol{a}}\;
\boldsymbol{P}_{h\perp} \cdot \hat{\boldsymbol{b}}} &= 
A_{UL}^{\boldsymbol{P}_{h\perp} \cdot \hat{\boldsymbol{a}}\;
\boldsymbol{P}_{h\perp} \cdot \hat{\boldsymbol{b}}}.
\end{align}
The $\sin \phi_h$ asymmetry can be rewritten as
\begin{equation} \begin{split} 
A_{UL}^{|\boldsymbol{P}_{h\perp}| \sin \phi_h} = 
\frac{2\sqrt{1-y}}{(1-y+y^2/2)\, f_1 D_1}\
\frac{M M_h}{Q} \bigg[ & \begin{aligned}[t] ( 2-y ) \bigg( & 
\frac{m}{M}z\ g_1 H_1^{\perp(1)} - \frac{M_h}{M} g_1 G^{\perp(1)} - 
xz\ h_L H_1^{\perp(1)} \\
& + h_{1L}^{\perp(1)} \tilde{H} - \frac{M}{M_h}xz\ f_L^{\perp(1)} D_1 \bigg)
\end{aligned} \\
&+ ( 1-y ) \bigg( xz h_{1} H_1^{\perp(1)} \bigg) \\
&-( 1-y+y^2/2 ) \bigg( \frac{M}{M_h}xz\ f_{1T}^{\perp(1)} D_1 \bigg)  \bigg]. 
\end{split} \end{equation}  
Neglecting quark masses, the asymmetry for jet production reads
\begin{equation}
A_{UL', j}^{|\boldsymbol{P}_{h\perp}| \sin \phi_h} =-\frac{M^2}{Q}\,
\bigg[
2 \sqrt{1-y}\ \frac{x\ f_{1T}^{\perp(1)}}{f_1} + 
\frac{2(2-y)\sqrt{1-y}}{(1-y+y^2/2)}\ \frac{f_L^{\perp(1)}}{f_1}
\bigg]
\end{equation}

The measurement of this asymmetry, at facilities where jet DIS can be
performed off polarized nucleons (e.g. eRHIC), would allow to determine the
size of the terms that contaminate the single-hadron $A_{UL}$
asymmetry. This asymmetry has been interpreted neglecting the contributions of the
Sivers function $f_{1 T}^{\perp}$ and of $f_L^\perp$ and $G^\perp$, leading to
predictions about the transverse spin asymmetry $A_{UT}$ that are not in good
agreement with preliminary data from the HERMES
collaboration~\cite{Seidl:2004}.
Finally, we point out {\em en passant} 
that  
jet DIS off {\em transversely} polarized nucleons  ($A_{UT,j}$ asymmetry) would be perhaps
the best way to pin down the Sivers function.

%%%%%%%%%%%%%%%%%%%%%%%%%%%%%%%%%%%%%%%%%%%%%%%%%%%%%%%%%%%%
\section{Conclusions}

In this paper we presented a complete study of 
the semi-inclusive DIS beam and target longitudinal spin
asymmetries, $A_{LU}$ and $A_{UL}$,  
up to subleading order in $1/Q$, including transverse momentum
dependent and T-odd distribution and fragmentation functions.

In order to be sure to include all contributions, we performed a new analysis
of the quark correlation functions, on the basis of what was
suggested in Ref.~\cite{Goeke:2003az}, where the necessity to
include the direction of the gauge link as an independent degree of freedom in
the decomposition of the correlation function was advocated. 
This revealed the existence of a new
distribution function never discussed before, which we named $g^\perp$, and
the analogous fragmentation function $G^\perp$. 
The new functions are T-odd, depend on
transverse momentum, are $1/Q$ suppressed and require no hadron polarization.
The very existence of these functions is related to the fundamental importance
of the gauge link in the definition of the correlation functions and in
particular to observable evidences of the light-cone direction the gauge link
runs along. 

Both functions turn out to contribute to the beam single spin asymmetry,
$A_{LU}$. The present description of such asymmetry~\cite{Gamberg:2003pz} 
is therefore incomplete.  
In particular, the term containing the function $g^\perp$ 
is the only one that can appear also in jet
semi-inclusive DIS, i.e., when the transverse momentum of the jet is observed,
instead of the transverse momentum of one hadron. Recent 
model calculations~\cite{Afanasev:2003ze,Metz:2004je} showed the occurrence of
nonzero beam longitudinal spin asymmetries in jet DIS. The connection between 
those model
calculations and our formalism has still to be carried out. However, they 
could possibly constitute a proof the
necessity of introducing the function $g^\perp$ and thereby corroborating the claims
of Ref.~\cite{Goeke:2003az}. An experimental check of a nonzero jet asymmetry
would be of
great importance, and could be done at ZEUS and H1, or at a new facility such eRHIC.

For what concerns the target longitudinal single spin asymmetry, $A_{UL}$, we
have found two extra terms compared to the existing literature, one containing
the function $G^\perp$ and one containing 
 the T-odd distribution function $f_L^\perp$, whose existence was already
 known but whose contribution to the $A_{UL}$ was so far neglected~\cite{Efremov:2003eq,Efremov:2003tf,Schweitzer:2003yr}. This
 finding has to be taken into account in analyses of the asymmetry, and could
 provide an explanation for the different behavior of the $A_{UL}$ and
  preliminary $A_{UT}$ asymmetries observed by the HERMES collaboration.

\begin{acknowledgments}
Discussions with D.~Boer and A.~Metz
are gratefully acknowledged. The work of A.~B. has been 
supported by the Alexander von Humboldt Foundation and BMBF. 
F.~P. thanks  the Institute
for Nuclear Theory at the University of Washington and the Department of Energy
for partial support of this work.
\end{acknowledgments}

%%%%%%%%%%%%%%%%%%%%%%%%%%%%%%%%%%%%%%%%%%%%%%%%%%%%%%%%%%%%%%%%%%%%%%%%%%%%%%%%
%%%%%

\bibliographystyle{apsrev}
\bibliography{mybiblio}

\end{document}